%% file: main.tex
\documentclass[conference]{IEEEtran}
\IEEEoverridecommandlockouts

\usepackage{cite}
\usepackage{amsmath,amssymb,amsfonts}
\usepackage{algorithmic}
\usepackage{graphicx}
\usepackage{textcomp}
\usepackage{xcolor}
\def\BibTeX{{\rm B\kern-.05em{\sc i\kern-.025em b}\kern-.08em
    T\kern-.1667em\lower.7ex\hbox{E}\kern-.125emX}}

\input{sections/00_preamble}

\begin{document}

\title{PackKV: Reducing KV Cache Memory Footprint through LLM-Aware Lossy Compression}

\author{%
\IEEEauthorblockN{Bo Jiang\IEEEauthorrefmark{1},
Taolue Yang\IEEEauthorrefmark{1},
Youyuan Liu\IEEEauthorrefmark{1},
Xubin He\IEEEauthorrefmark{1},
Sheng Di\IEEEauthorrefmark{2},
Sian Jin\IEEEauthorrefmark{1}}
\IEEEauthorblockA{\IEEEauthorrefmark{1}\textit{Temple University}, Philadelphia, USA}
\IEEEauthorblockA{\IEEEauthorrefmark{2}\textit{Argonne National Laboratory}, Lemont, USA}
\IEEEauthorblockA{\{jiang.bo, taolue.yang, youyuan.liu, xubin.he, sian.jin\}@temple.edu; sdi1@anl.gov}
}


\maketitle

\input{sections/01_abstract}
\input{sections/02_introduction}
\input{sections/03_background}

\input{sections/04_methodology}
\input{sections/05_experiment}
\input{sections/06_conclusion}

\clearpage

\bibliographystyle{IEEEtran}
\bibliography{references}

\end{document}

%% file: sections/00_preamble.tex
\usepackage{subcaption}
\usepackage{booktabs}
\usepackage{multirow}
\usepackage{algorithm}
\usepackage{hyperref}
\usepackage[numbers,sort&compress]{natbib}
\usepackage[most]{tcolorbox}

\newcommand{\cdg}{\textcolor[RGB]{0,150,0}}
\usepackage{tikz}

\newcommand{\blackcircled}[1]{%
  \tikz[baseline=(char.base)]{
    \node[shape=circle, fill=black, text=white, inner sep=1pt] (char) {#1};
  }%
}
\definecolor{RCmt}{HTML}{E64A19} 
\newcommand{\RC}[1]{\hfill{\color{RCmt}\footnotesize\bfseries$\blacktriangleright$~#1}}

\definecolor{kvAccent}{HTML}{2F6690} 
\newcommand{\smx}{\operatorname{softmax}}
\newcommand{\Att}{\operatorname{Att}}
\tcbset{
  kvbox/.style={
    enhanced, breakable,
    colback=white, colframe=kvAccent!35, boxrule=0.4pt, arc=2pt,
    colbacktitle=kvAccent!8, coltitle=black, fonttitle=\bfseries,
    borderline west={2.2pt}{0pt}{kvAccent},
    left=7pt, right=7pt, top=6pt, bottom=6pt
  }
}

%% file: sections/01_abstract.tex
\begin{abstract}
Transformer-based large language models (LLMs) have demonstrated remarkable potential across a wide range of practical applications. However, long-context inference remains a significant challenge due to the substantial memory requirements of the key-value (KV) cache, which can scale to several gigabytes as sequence length and batch size increase. In this paper, we present \textbf{PackKV}, a generic and efficient KV cache management framework optimized for long-context generation.
PackKV introduces novel lossy compression techniques specifically tailored to the characteristics of KV cache data, featuring a careful co-design of compression algorithms and system architecture. Our approach is compatible with the dynamically growing nature of the KV cache while preserving high computational efficiency. Experimental results show that, under the same and minimum accuracy drop as state-of-the-art quantization methods, PackKV achieves, on average, \textbf{153.2}\% higher memory reduction rate for the K cache and \textbf{179.6}\% for the V cache. Furthermore, PackKV delivers extremely high execution throughput, effectively eliminating decompression overhead and accelerating the matrix-vector multiplication operation. Specifically, PackKV achieves an average throughput improvement of \textbf{75.7}\% for K and \textbf{171.7}\% for V across A100 and RTX Pro 6000 GPUs, compared to cuBLAS matrix-vector multiplication kernels, while demanding less GPU memory bandwidth. Code available on \url{https://github.com/BoJiang03/PackKV}
\end{abstract}


%% file: sections/02_introduction.tex
\section{Introduction}

\textcolor{black}{Transformer-based large language models (LLMs) have revolutionized natural language processing, enabling breakthroughs in diverse tasks~\cite{brown2020language, taylor2022galactica}.}
The self-attention mechanism allows models to capture long-range dependencies and contextual information. 
However, these capabilities come at a significant computational and memory cost during inference with long input contexts, \textcolor{black}{where the memory footprint of the key-value (KV) cache becomes a major bottleneck~\cite{kwon2023efficient, pope2023efficiently}.}

This growing footprint severely constrains the inference performance, \textcolor{black}{limiting the achievable context length, reducing batch size, or impeding the deployment of LLMs on memory-constrained hardware~\cite{kwon2023efficient, liu2023scissorhands}.} The KV cache stores intermediate key and value tensors for each token processed by the model and is reused during subsequent decoding steps to avoid redundant computation. As sequence length and batch size increase, the cache size grows linearly and can consume a substantial portion of GPU memory, \textcolor{black}{sometimes exceeding the memory footprint of the model weights themselves~\cite{liu2023scissorhands}.} For example, LLaMA2-30B inference with a context length of 32,000 and a batch size of 8 can produce over 100 GB of KV cache, surpassing the model size itself of 60 GB in float16.

To address these challenges, recent studies mainly leverage three approaches to reduce KV cache size: quantization, pruning, and GPU-CPU migration. 
\textcolor{black}{Quantization-based methods~\cite{liu2024kivi, hooper2024kvquant}}, such as KIVI, aim to carefully design quantization strategies that minimize the impact on model accuracy. 
However, these techniques often achieve modest compression ratios unless combined with additional encoding, which introduces overhead and limits their applicability in latency-sensitive LLM inference. 
\textcolor{black}{Pruning-based methods~\cite{wang2021spatten, zhang2024q}}, such as Q-Hitter, selectively discard KV pairs that are predicted to be unimportant for future decoding. 
While effective in some cases, these methods suffer from unpredictable attention callbacks, leading to either costly KV recomputation or substantial accuracy degradation. 
\textcolor{black}{GPU-CPU migration is a traditional approach for handling memory overflows, offloading KV cache data to CPU memory~\cite{sheng2023flexgen}.} Although this mitigates GPU memory pressure, it significantly degrades inference performance due to data transfer latency and complex scheduling overhead.

In this paper, we present PackKV, a high-performance, LLM-aware lossy compression framework tailored for KV cache optimization during inference. PackKV integrates error-controlled quantization, a KV-cache–specific lossless compression scheme, and cache-resident decompression to achieve substantial memory savings while improving computational efficiency. By co-designing the compression pipeline with system-level execution, PackKV enables decompressed data to be consumed in situ within GPU shared memory and registers, thereby eliminating global memory writebacks and maximizing throughput. This high-throughput design allows PackKV to be deployed in LLM inference with negligible overhead, and in most cases, to even accelerate inference computation.
Importantly, our approach is orthogonal to existing pruning methods and GPU–CPU migration strategies.

Our key contributions are as follows:
\begin{itemize}
    \item \textbf{Novel compression pipeline design:} We introduce \textbf{PackKV}, the first \textit{LLM-aware lossy compression framework} that integrates \textit{quantization}, \textit{encode-aware repacking}, and \textit{bit-packing encoding}. PackKV achieves \textbf{153.2\%} and \textbf{179.6\%} higher compression ratios for \textit{K} and \textit{V} caches, respectively, compared with state-of-the-art quantization methods with matched benchmark accuracy.

    \item \textbf{Computation-aware decompression integration:} We propose a \textit{co-design strategy} that embeds decompression directly into \textit{matrix–vector multiplication kernels}, thereby eliminating global memory writebacks of decompressed data and redundant global memory reads during computation. This design yields \textbf{75.6\%}(K) and \textbf{171.6\%}(V) throughput improvements, compared to cuBLAS.

    \item \textbf{Theoretical foundation and practical algorithms:} We formulate the \textit{KV cache reordering problem} as an optimization variant of the \textit{set partition problem}, providing both an \textit{greedy solution} and an \textit{efficient median-based algorithm} with lower time complexity. Furthermore, we utilize the \textit{permutation invariance} of attention computation, enabling \textit{lossless compression ratio optimization}.

    \item \textbf{Comprehensive system implementation:} We develop a \textit{complete GPU-based framework} featuring a \textit{block-independent compression format} that supports \textit{seamless appending} for dynamic KV cache growth, \textit{single-kernel decompression} across multiple blocks, and \textit{perfect multi-GPU scalability} with near zero interference.

    \item \textbf{Extensive experimental validation:} Evaluations across \textbf{six LLM models} and \textbf{six benchmark datasets} demonstrate that PackKV consistently \textit{outperforms state-of-the-art quantization methods} in \textit{compression ratio}, and \textit{improves throughput} compared to cuBLAS kernels.
\end{itemize}

The rest of this paper is organized as follows: Section \ref{sec:background} provides a background on LLM inference, lossy compression, and GPU architecture. Section \ref{sec:design} presents the design of our PackKV framework. Section \ref{sec:evaluation} details our experimental evaluation and analysis. Lastly, Section \ref{sec:conclusion} concludes the paper and future research directions.

%% file: sections/03_background.tex
\section{Background and Motivation}
\label{sec:background}

\subsection{Large Language Model Inference}

LLMs have become foundational in natural language processing, demonstrating remarkable capabilities in text generation~\cite{achiam2023gpt}. The inference process involves auto-regressive decoding, where the model generates tokens one by one, with each new token depending on previously generated ones and the initial prompt~\cite{radford2018improving, vaswani2017attention}.

To enable efficient attention to preceding tokens without recomputing states at every generation step, Transformer models maintain a Key–Value (KV) cache~\cite{shazeer2019fast}. The core component enabling this process is the attention mechanism~\cite{vaswani2017attention}. During inference, Key (K) and Value (V) tensors are computed for each token within self-attention layers; these K and V tensors are then reused to process future tokens and collectively form the KV cache~\cite{shazeer2019fast}. The KV cache is organized as $[context\_len, head\_num, head\_dim]$, where $context\_len$ is the sequence length, $head\_num$ is the number of attention heads, and $head\_dim$ is the head dimension.


\begin{figure}[t]
  \centering
  \begin{subfigure}[b]{0.45\textwidth}
        \centering
        \includegraphics[width=\textwidth]{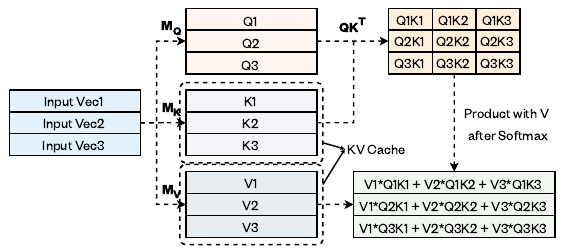}
        \caption{KV cache in prefill stage(assuming 3 token input).}
        \label{fig:kv_cache_prefill}
    \end{subfigure}
    \hfill
    \begin{subfigure}[b]{0.45\textwidth}
        \centering
        \includegraphics[width=\textwidth]{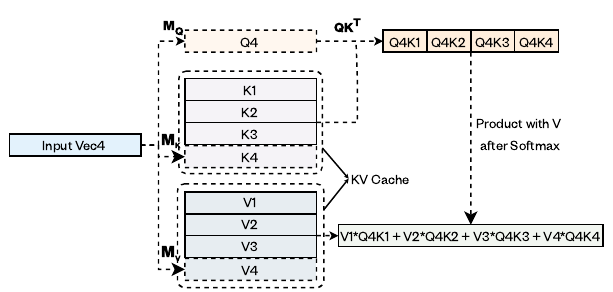}
        \caption{KV cache in decode stage(generating one token).}
        \label{fig:kv_cache_decode}
    \end{subfigure}
    \caption{KV Cache behavior during LLM Inference, \(M_{Q,K,V}\) is the mapping matrix to \(K,Q,V\) vector. Each \(Q\) vector performs dot product with every \(K\) vector to generate a \textit{weight} after softmax operation for each \(V\) vector. All the \(V\) vectors multiplied with their \textit{weights} and aggregate to \textit{Attn Output}.}
    \vspace{-0.5cm}
\end{figure}

KV cache participates in two stages during inference~\cite{agrawal2023sarathi}.
\textbf{Prefill Stage:}  Initial prompt processing generates the initial KV cache (Figure~\ref{fig:kv_cache_prefill}).
\textbf{Decode Stage:}  Auto-regressive token generation where new KV vectors are appended to the cache and the entire cache is used for attention computation via matrix-vector multiplications (Figure~\ref{fig:kv_cache_decode}).

By rapidly inflating GPU memory usage, the KV cache often becomes the bottleneck that constrains sequence length and batch size, degrading performance or even preventing inference on memory-constrained hardware. Although it enables efficient inference, the cache grows linearly with sequence length and batch size and can consume more GPU memory than model weights \cite{kwon2023efficient, liu2023scissorhands}. Existing remedies include KV-cache quantization \cite{liu2024kivi, hooper2024kvquant}, which provides limited reduction at near-original accuracy, and GPU--CPU offloading \cite{sheng2023flexgen}, which lowers GPU memory needs at the cost of substantial performance overhead.

Memory bandwidth, rather than compute, becomes the primary bottleneck in autoregressive decoding as the KV cache comes to dominate both memory footprint and memory traffic. Decoding operations reduce to matrix–vector multiplications that dominate runtime (93.71\% of GPU kernel time with a 100k-token context in our Code Llama experiments) and are memory-bound due to low arithmetic intensity and poor data reuse. During decoding, both model weights and the entire KV cache must be read once from GPU global memory; as context length increases, the KV cache dominates memory consumption—for example, CodeLlama 2–7B with a 100k-token context requires 50~GB for the KV cache versus 14~GB for model weights—so the cache exceeds 78\% of the total memory footprint and creates significant bandwidth bottlenecks.

\subsection{Data Reduction in LLMs}


\begin{figure*}[]
    \centering
   \includegraphics[width=\linewidth]{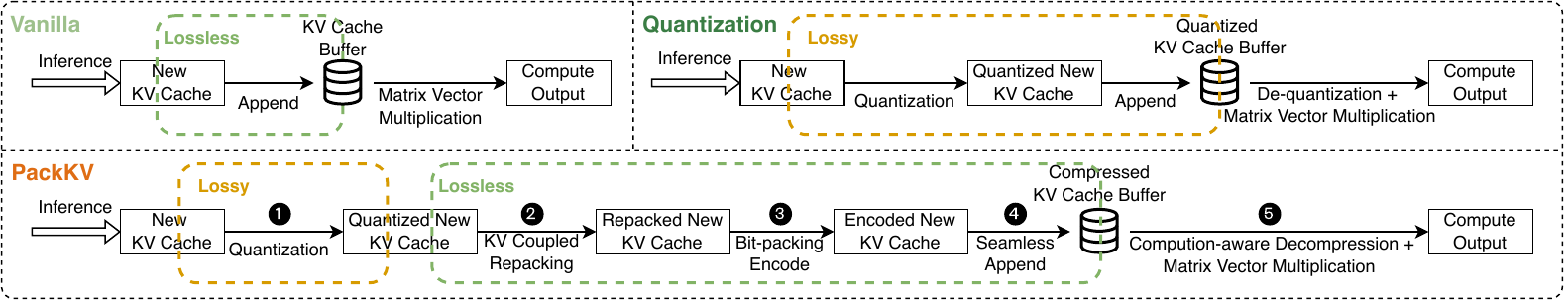}
    \caption{Overview of PackKV. Compared with vanilla and previous quantization solutions. (1), We rearrange the order of quantized kv cache vector to benefit the following lossless bit-packing encode. (2). Bit-packing the quantized kv cache integers in channel direction to achieve extra memory saving. And to eliminate the decompress overhead we developed a Computation-aware Decompression + matrix-vector Multiplication method.} 
    \label{fig:overview}
    \vspace{-0.3cm}
\end{figure*}

In large neural networks, particularly LLMs, \textcolor{black}{lossy compression can be applied to large data structures like activations or caches to mitigate memory bottlenecks~\cite{han2016deep_compression,jacob2018quantization, banner2019post,frantar2022gptq, jin2019deepsz, jin2021comet}}. 
For example, quantization is a form of lossy compression that reduces the precision of numerical data by mapping continuous or high-precision values (e.g., float32) to a smaller, discrete set of values such as integers.
In fact, this inherently reduces the entropy of the data, making it smaller and often more amenable to lossless encoding.
\textcolor{black}{In image, video, and scientific lossy compression, entropy encoding and spatial encoding are commonly added to further reduce data size~\cite{wallace1992jpeg,taubman2002embedded,lindstrom2014fixed, ohm2012comparison, di2016fast, tao2017significantly, sz18}}.
However, the low throughput of these post-quantization encoders makes them impractical for real-world LLM inference, where latency is critical. 
For example, \textcolor{black}{CacheGen~\cite{liu2024cachegen} integrates such encoders into the KV Cache quantization pipeline and achieves throughput below 1 GB/s, which is an improvement over network transmission, but insufficient compared to the GPU-CPU transmission and GPU memory bandwidth.}

Another challenge in applying lossy compression is to manage the trade-off between the desired compression ratio and the potential impact on model accuracy.
Applying quantization requires careful consideration of the data's characteristics and the specific downstream task. Different quantization strategies, such as varying granularity (e.g., token/channel/block-wise) or using adaptive error bounds, can be employed to control the information loss and minimize accuracy degradation~\cite{liu2024kivi}. 

\subsection{GPU architecture}




\textcolor{black}{The GPU memory hierarchy plays a crucial role in performance optimization~\cite{mei2016dissecting}.}
Global memory provides the largest capacity with high latency, making efficient access patterns essential. 
\textcolor{black}{For example, coalesced memory access can significantly improve bandwidth utilization.}
While shared memory in each SM is a programmer-managed cache shared among threads within a block.


GPU execution follows a Single Instruction, Multiple Thread (SIMT) model, where threads within a warp (typically 32 threads) execute the same instruction in lockstep. This model creates challenges when threads need to follow different execution paths, which is also called branch divergence that forces serialization and reduces performance. 
To coordination among threads, GPUs provide atomic operations that ensure read-modify-write sequences complete without interference, essential for managing shared resources or global indices across thread blocks.


%% file: sections/04_methodology.tex
\section{Design Methodology}
\label{sec:design}

This section presents our KV cache compression framework--\textbf{PackKV} for LLM inference. 
Figure~\ref{fig:overview} shows our five-step pipeline: \blackcircled{1} \textbf{Quantization}, \blackcircled{2} \textbf{Encode-aware Repacking}, \blackcircled{3} \textbf{Bit-packing}, \blackcircled{4} \textbf{Seamless Appending}, and \blackcircled{5} \textbf{Computation-aware Decompression}.

Our framework processes KV cache in blocks stored in a fixed-size buffer. In step \blackcircled{1}, each token's KV cache is quantized using a shared quantization scale. Step \blackcircled{2} reorders the quantized vectors to maximize compression efficiency while keeping K-V pairs together. We demonstrate that attention computation is order-invariant, this reordering incurs no decompression overhead. Step \blackcircled{3} applies lightweight bit-packing encoding with carefully managed metadata. Step \blackcircled{4} appends compressed blocks using a format that enables single-kernel decompression, eliminating multiple kernel launch overhead. Finally, step \blackcircled{5} integrates decompression with matrix-vector multiplication, decompressing data directly into GPU registers to reduce memory access and accelerate the primary bottleneck of LLM inference.

\subsection{Algorithm Design}
\label{sec:algorithm_design}


Our compression pipeline consists of two main components: \textbf{Lossy Compression via Quantization} and \textbf{Lossless Compression via encode-aware repacking and bit-packing}. In the quantization step, high-precision values are mapped to discrete integers, reducing data entropy. For lossless compression, we first apply \emph{encode-aware repacking}: quantized KV-cache vectors are reorganized to groups, enabling the subsequent encoder to use a shorter per-pack encoding length and achieve higher compression ratios. We then encode these packs using a lightweight \emph{bit-packing} scheme that, for each pack, selects the shortest encoding length (bit-width) required to represent all elements and packs them accordingly.

\textbf{Why quantization followed by lossless compression?} This design keeps the introduced error simple. By applying only lossless compression after quantization, our method maintains theoretically the same accuracy as quantization-only solutions, making error analysis straightforward.

\begin{figure}[]
    \centering
    \vspace{1.5em}
    \includegraphics[width=\linewidth]{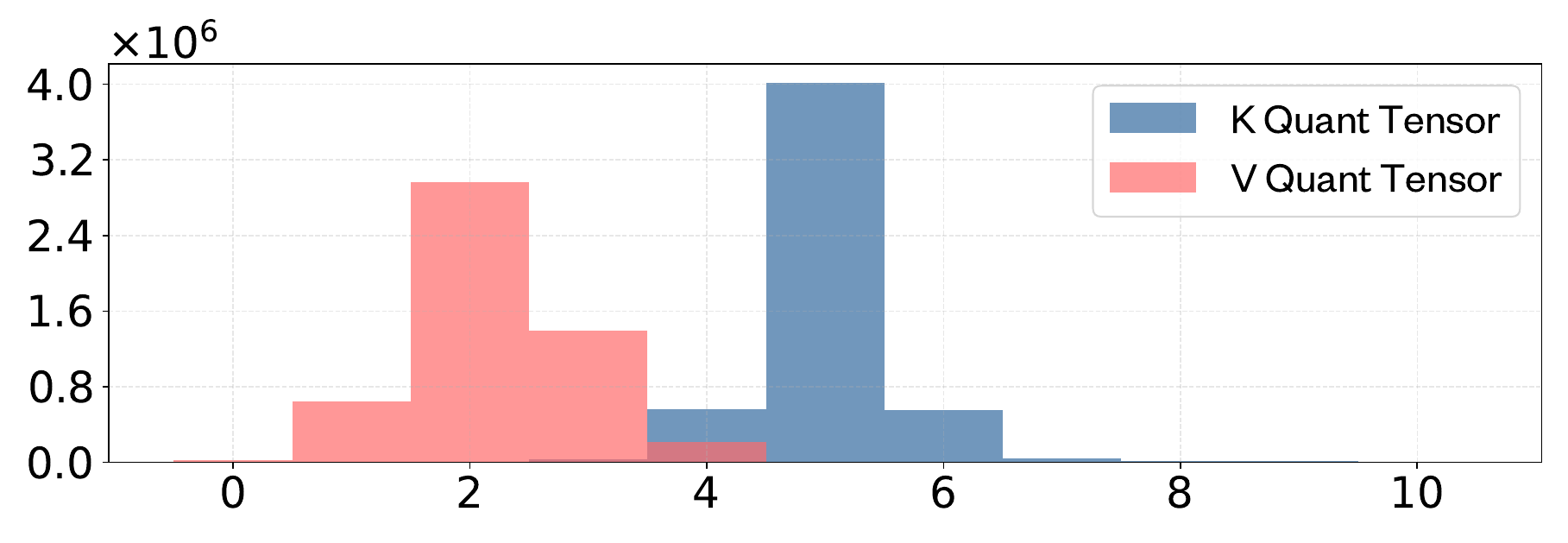}
    \caption{Histogram of K and V quantized integer tensor, x-axis is quantized integer value, y-axis is the frequency of the value.}
    \label{fig:histogram}
\end{figure}

\begin{figure}[]
  \centering
  \includegraphics[width=0.74\linewidth]{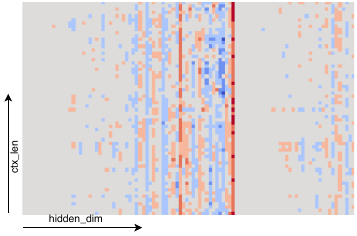}
  \caption{A visualization of a sampled K after quantization to integer, horizontal direction is hidden\_dim and vertical is context\_len (channel) direction.}
  \label{fig:k_heatmap}
  \vspace{-1em}
\end{figure}

\textbf{Why encode-aware repacking and bit-packing?} We target a large compression–decompression asymmetry. On Llama2-13B (40 layers, 5120 hidden, FP16/BF16) at 50 tokens/s, KV-cache generation is only 39.06~MB/s—modest for modern compressors—whereas decoding must restore the entire KV cache for every new token; with thousand-token contexts, required decompression throughput is 3–5 orders of magnitude higher compared to compression. Thus, we need an ultra-light decompressor that exploits the low-entropy distributions of quantized data (Figure~\ref{fig:histogram}) and repeating patterns (Figure~\ref{fig:k_heatmap}), and that can be tightly fused with computation to cut memory traffic or even accelerate memory-bound kernels. Bit-packing best matches these constraints: it stores integers with minimal bits, excels on low-entropy, repetitive data, and we fuse it with matrix–vector operations. Encode-aware repacking further boosts compression by reordering KV vectors (leveraging the Permutation Invariance of KV Cache, introduced in \ref{proof:permutation_invariance}) to increase per-pack homogeneity, improving the achievable encoding length without adding decompression overhead.

\subsection{Lossy Compression Design} \label{sec3.1}

PackKV consists of three key steps:
(1) \textbf{Buffering and Blocking:} Temporarily stores KV cache vectors and feeds complete blocks into the compressor.
(2) \textbf{Encode-aware Repacking and Bit-packing:} Reorganize KV cache vectors to improve encoding efficiency, then applies bit-packing with adaptive storage formats for K and V.
(3) \textbf{Seamless Appending:} Uses block-independent format to decouple compression/decompression, enabling single-kernel decompression.


\subsubsection{Buffering and Blocking}

To handle the dynamically growing KV cache, we use a fixed-size buffer of shape [\textit{max\_buffer\_size}, \textit{head\_num}, \textit{head\_dim}]. When the buffer overflows, it is truncated and partitioned into 2D blocks for compression(\textit{head\_num}, \textit{head\_dim} flattened).

For K cache, we use 2D blocks with \textit{head\_dim} as one dimension to ensure entire dot product vectors are contained within single GPU threads block, eliminating inter-block data exchange in computation-aware decompression. For V cache, we swap the block dimensions since dot products occur along the \textit{context\_len} dimension. This design optimizes both decompression efficiency and computational performance.

\subsubsection{Quantization}

Insitu quantization is the only lossy step in our compression pipeline.

\begin{figure}[]
  \centering
  \includegraphics[width=\linewidth]{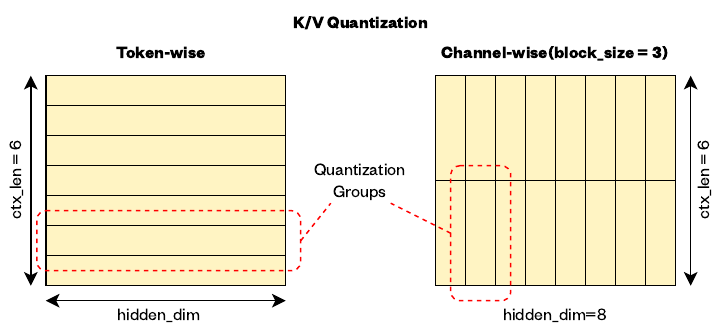}
  \caption{KV quantization granularities based on KV cache dimensions: token-wise and channel-wise quantization.}
  \label{fig:quantization}
  \vspace{-1em}
\end{figure}

Several quantization granularities can be applied to the KV cache tensor (Figure~\ref{fig:quantization}). KIVI~\cite{liu2024kivi} uses fine-grained channel-wise quantization for K cache and token-wise for V cache, exploiting channel correlations (i.e., column-wise in Figure~\ref{fig:k_heatmap}). However, quantization-only methods have limited flexibility as bit-widths must be integers (2, 3, or 4 bits). Higher compression ratios require reducing bit-width, but this demands smaller quantization units to maintain model accuracy, increasing metadata overhead. For KIVI's 2-bit K quantization with 64-element blocks which is recommended by KIVI (even 128 is not recommended), the actual compression ratio is only 6.4 (not 8) due to FP16 metadata. If bit-width falls back to 3 bits, compression drops to 4.57 which is close to that of coarse-grained 4-bit quantization. This reveals \textbf{quantization's dilemma}: Using finer quantization granularity to reach lower bit widths inflates metadata, capping overall compression.

We adopt \textbf{token-wise quantization for both K and V caches}, relying on subsequent lossless compression to capture data patterns and entropy characteristics. 

\subsubsection{Encode-aware Repacking and Bit-packing Encoding}

We adopt bit-packing as our encoding method, allocating only the exact number of bits needed for each quantized integer in each encode group. We define a group of integers as a \emph{pack}, determine its value range, and calculate the required bits as $\lceil \log_2(\text{range} + 1) \rceil$, where $\text{range} = \text{max} - \text{min}$. Two key configurations determine this process: 
\begin{itemize} 
    \item \textbf{Pack size}: number of integers per pack 
    \item \textbf{Encoding direction}: for encoding 2D quantized data.
\end{itemize}

Pack size depends on data characteristics and is determined empirically in our experiments. Encoding direction is influenced by data correlations (strong mutual information in K cache channel dimension) and the permutation invariance of KV cache vectors. 

\label{proof:permutation_invariance}
\begin{tcolorbox}[kvbox, title=Permutation Invariance of KV Cache]
\small
\textbf{Summary.} Jointly permuting the rows of K and V does not change the attention output (after positional embedding).

\textbf{Mathematical Statement.}
For $q\in\mathbb{R}^{d_k}$, $K\in\mathbb{R}^{L\times d_k}$, $V\in\mathbb{R}^{L\times d_v}$,
and a permutation matrix $P\in\mathbb{R}^{L\times L}$, define
\[
\Att(q,K,V)
= \smx\!\Big(\tfrac{1}{\sqrt{d_k}}\,Kq\Big)^{\!\top} V.
\]
Then
\[
\Att(q,PK,PV)=\Att(q,K,V).
\]

\textbf{Proof (sketch).}
Let $s=\tfrac{1}{\sqrt{d_k}}\,Kq$, $\alpha=\smx(s)$ and note $\smx(Pz)=P\,\smx(z)$.
Hence
\[
\begin{aligned}
\Att(q,PK,PV)
&= \smx\!\Big(\tfrac{1}{\sqrt{d_k}}\,PKq\Big)^{\!\top} (PV) \\
&= (P\alpha)^\top(PV) \\
&= \alpha^\top V \;=\; \Att(q,K,V).
\end{aligned}
\]

\textbf{Note.}
Prefill uses a fixed lower-triangular mask (not invariant to column permutations), so a global permutation breaks prefill. Decoding has a single active row and is effectively unmasked, so permutations are valid. We validated this on the CoQA benchmark, where randomly permuting KV cache order during decoding produced unchanged results.
\end{tcolorbox}

\vspace{-2em}

This permutation invariance allows us to reorder vectors along the block\_size dimension to optimize bit-packing effectiveness. After determining the pack size and encoding direction, the key challenge is determining the optimal reordering strategy to maximize compression ratio:

\textbf{Problem Formalization: }
\label{para:problem_formalization}Let $A$ be a finite set with $|A| = n$. A \emph{legal partition} of $A$ is a collection of disjoint subsets $\mathcal{P} = \{S_1, S_2, \ldots, S_g\}$ such that $S_i \subseteq A$, $S_i \cap S_j = \emptyset$ for all $i \neq j$, and $\bigcup_{i=1}^g S_i = A$.

Given a real-valued function $f$ defined on subsets of $A$, the objective is to find a partition $\mathcal{P}$ that minimizes the total cost:
\vspace{-5em}
\[
\min_{\mathcal{P}} \sum_{i=1}^g f(S_i)
\vspace{-4em}
\]
\textbf{Problem Specification: }In our scenario, $A$ represents quantized KV cache vectors within a block, where each $\mathbf{a} \in A$ is a $d$-dimensional non-negative integer vector ($\mathbf{a} \in \mathbb{Z}_+^d$, typically $d = 5120$ for Llama2-13B).

Each group $S_i$ has size $k$ (pack size), with $n$ vectors partitioned into $g = n / k$ groups. The cost function $f(S_i)$ is the total bits required to encode group $S_i$ using bit-packing: subtracting per-dimension minimums, determining bits needed for value ranges, and summing storage costs including metadata overhead.

\textbf{Problem Analysis: }The formalized problem reduces to an optimization version of the set partition problem, which is NP-hard \cite{rasmussen2011optimisation}. Given this complexity, we propose a greedy algorithm \ref{alg:greedy_repacking} that incrementally constructs packs by selecting vectors that locally minimize incremental cost:

\begin{algorithm}[H]
\caption{Greedy Repacking for Bit-Packing}
\label{alg:greedy_repacking}
\begin{algorithmic}[1]
\REQUIRE $X=\{x_i\}_{i=1}^N$ with $x_i\in\mathbb{Z}^D$; capacity $k\in\mathbb{N}$; cost $C:2^{\{1,\dots,N\}}\to\mathbb{R}_{\ge 0}$ \RC{inputs and pack cost}
\ENSURE repacked order given by concatenation of emitted packs \RC{repacked order}
\STATE $\mathcal{R}\gets \{1,\dots,N\}$ \RC{unassigned indices}
\WHILE{$\mathcal{R}\neq\emptyset$}
    \STATE $c \gets \frac{1}{|\mathcal{R}|}\sum_{i\in\mathcal{R}} x_i$ \RC{centroid of remaining vectors}
    \STATE $s \in \arg\min_{i\in\mathcal{R}} \|x_i - c\|_2$;\; $P\gets\{s\}$;\; $\mathcal{R}\gets \mathcal{R}\setminus\{s\}$ \RC{seed = closest to centroid}
    \WHILE{$|P|<k$ and $\mathcal{R}\neq\emptyset$}
        \STATE $j^\star \in \arg\min_{j\in\mathcal{R}} \big[C(P\cup\{j\})-C(P)\big]$ \RC{least marginal cost}
        \STATE $P \gets P\cup\{j^\star\}$;\; $\mathcal{R}\gets \mathcal{R}\setminus\{j^\star\}$ \RC{commit $j^\star$}
    \ENDWHILE
    \STATE emit $P$ \RC{append pack to output stream}
\ENDWHILE
\end{algorithmic}
\end{algorithm}

\vspace{-1em}

\begin{figure}[htbp]
  \centering
  \includegraphics[width=0.8\linewidth]{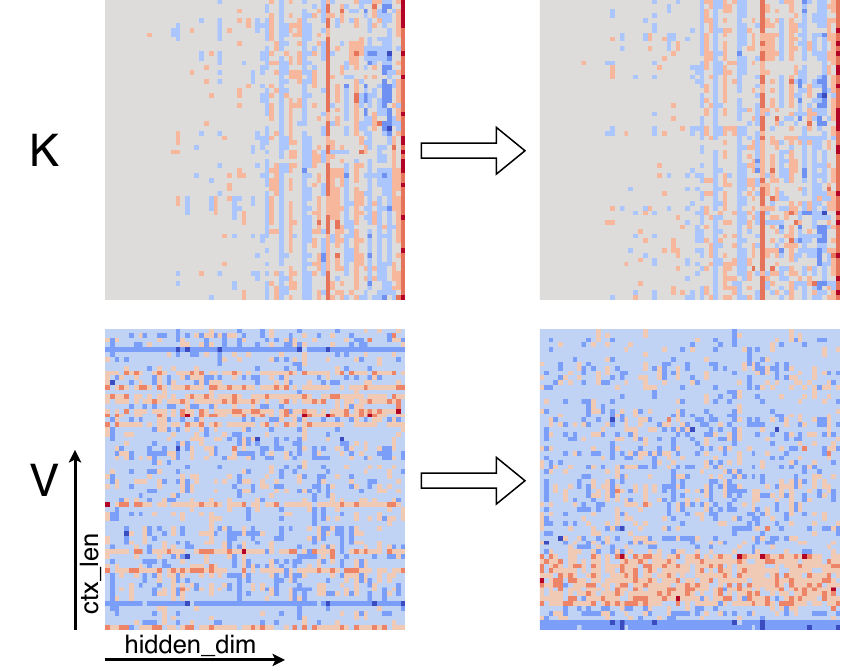}
  \caption{Before and after greedy repacking.}
  \label{fig:before_after_repacking}
  \vspace{-1em}
\end{figure}

The greedy algorithm forms $N/k$ packs, and $O(N)$ candidates for each of $k$ positions per pack with $O(D)$ cost evaluation per candidate. The overall time complexity is $O(N^2 D)$.

\textbf{Faster Repacking with V Median: }\label{sec:sec_v_median}The $O(N^2 D)$ greedy complexity is prohibitive for large blocks. Figure~\ref{fig:before_after_repacking} shows that greedy repacking groups vectors with similar V median values. We propose \emph{V Median Repacking}, which sorts vectors by their V part median, achieving $O(ND)$ complexity with radix sort while maintaining good compression effectiveness.




\subsubsection{Seamless Appending}

After compression, each 2D block is encoded as a contiguous bit chunk and assigned a unique index identifying its horizontal position (attention head) and vertical position (the token range) in the KV cache. Because blocks are compressed independently, newly produced blocks can be \textbf{seamlessly appended} to the end of the existing compressed buffer.
This enables efficient storage and incremental management of the cache over time, which typical lossy compressors do not support.

Frequent decompression GPU kernel launches incur high latency and cap throughput. For LLaMA2-13B with a 32K context, block size 64, and 512 separate launches, a $5~\mu s$ launch time limits throughput to about 131~GB/s, slowing attention computation. We therefore use \textbf{a single decompress+compute kernel} with many parallel GPU blocks to decode concurrently, minimizing launch overhead. This single-launch design relies on our flexible appendable format; without it, a single kernel implementation would not be feasible.

\subsection{Computation-Aware Decompression Design}
\label{sec:decompression}

Compression and decompression demands are highly imbalanced due to auto-regressive LLM generation.
As aforementioned in Section~\ref{sec:algorithm_design}, 
compression is applied only to newly generated tokens, whereas all cached tokens must be decompressed at least once to generate each new token. 
Without computation-aware decompression, decompression is serialized with subsequent K/V computations, and K/V kernels alone can sustain at 500+ GB/s based on our evaluation in Section~\ref{sec:evaluation}, which is far higher than existing state-of-the-art GPU lossy compressors (i.e., when compress fp16 data), 
in which case decompression would dominate end-to-end runtime, making any current standalone lossy compressor impractical for this scenario.

To improve decompression efficiency, we integrate decompression directly into matrix-vector multiplication. Since SOTA lossy compressor throughput approaches bandwidth limits and matrix–vector multiplication is a memory-bound operation, this integration brings two main benefits: (1) eliminates writing decompressed data back to global memory—a significant overhead given that decompressed data is several times larger than compressed data; and (2) reduces global memory transfers when loading matrices into shared memory or registers.

However, decompression and computation data orders differ, making integration challenging. Additionally, K and V cache dot products are orthogonal, and we use the same encoding method for both, further complicating decompression-computation integration. To address these challenges, we redesign the compressed data format to optimize data loading, and modify the parallelization strategy for decompression as well as the data reduction scheme in matrix-vector multiplication, making integration feasible.

\subsubsection{K Compressed Format and Computation Integration}~ 

\begin{figure}[]
  \centering
  \includegraphics[width=\linewidth]{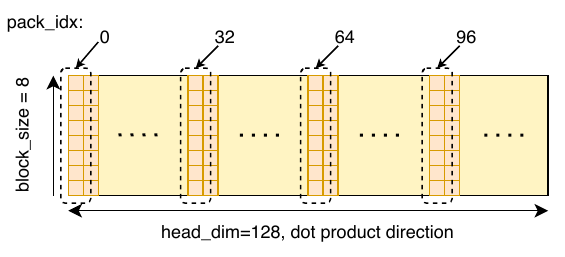}
  \vspace{-1em}
  \caption{A 2D block of K quantized integer tensors to be decoded by single GPU threads block.}
  \label{fig:k_quant_integer_tensor}
  \vspace{-1em}
\end{figure}


\begin{figure}[]
  \centering
  \vspace{1em}
  \includegraphics[width=\linewidth]{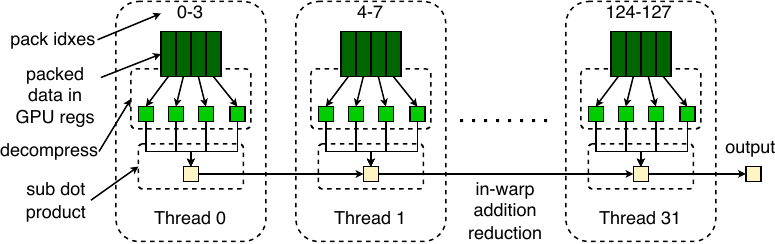}
  \caption{K decompress + dot product.}
  \label{fig:k_dot_product}
  \vspace{-1em}
\end{figure}

\textbf{Dot product thread assignment:} Matrix-vector multiplication consists of independent dot products. For GPU implementation, a key consideration is thread assignment per dot product. Figure~\ref{fig:k_quant_integer_tensor} shows our simplified example with block size 8, pack size 8, and one attention head.

For K Cache, bit-packing direction (ctx\_len) is orthogonal to dot product direction (head\_dim). Single-thread-per-dot-product requires multiple memory loads per multiply-accumulate operation, using only a few bits while discarding the rest, causing poor data locality. Therefore, we assign a warp to process multiple K dot products sequentially. With head\_dim=128 (typical for LLMs), each thread loads 4 encoded packs, decompresses 2 groups of 4 data points (half2, CUDA vector type packing two half-precision floats for efficient paired operations), and performs dot products. After completion, half2 warp-level sum reduction completes the two head\_dim vector operations shown in Figure~\ref{fig:k_dot_product}.

\begin{figure}[htbp]
  \centering
  \includegraphics[width=\linewidth]{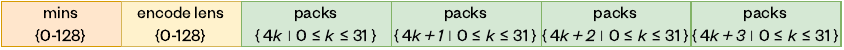}
  \caption{K compressed data format.}
  \label{fig:k_format}
\end{figure}

\textbf{Packs storage format:} To avoid bank conflicts when loading four packs into registers, we store packs 0, 4, 8, ..., 124 contiguously, then packs 1, 5, 9, ..., 125, and so on. Encoded lengths and minimum values are stored in the beginning of pack data, shown in Figure~\ref{fig:k_format}.

\subsubsection{V Compressed Format and Computation Integration}~

\begin{figure}[htbp]
  \centering
  \includegraphics[width=\linewidth]{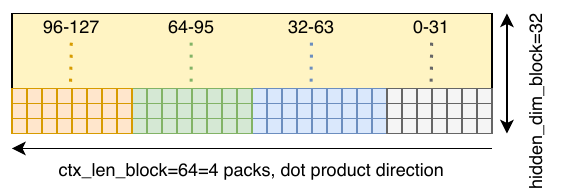}
  \caption{A 2D block of V quantized integer tensors to be decoded by single GPU threads block.}
  \label{fig:v_quant_integer_tensor}
\end{figure}


\begin{figure}[htbp]
  \centering
  \includegraphics[width=\linewidth]{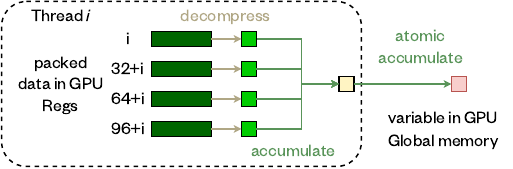}
  \caption{V decompress + dot product.}
  \label{fig:v_dot_product}
\end{figure}

\begin{figure}[htbp]
  \centering
  \vspace{-0.5em}
  \includegraphics[width=\linewidth]{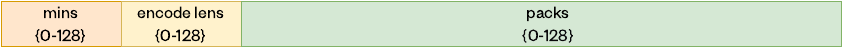}
  \caption{V compressed date format.}
  \label{fig:v_format}
  \vspace{-1em}
\end{figure}

\textbf{Dot product thread assignment:} One example V block for one GPU threads block is shown in Figure \ref{fig:v_quant_integer_tensor}. Unlike K Cache, V Cache dot product vector length is dynamic (1 to millions), making neither one-warp-per-dot-product nor one-thread-per-dot-product suitable. We assign each thread to compute a dot product segment, accumulating partial results using atomic add operations with float32 precision. Since dot product direction aligns with bit-packing encoding direction, each thread efficiently loads its pack and compute two dot products using half2 operations, shown in Figure~\ref{fig:v_dot_product}.

\textbf{Packs storage format:} We store pack data for indices 0-127 contiguously. No bank conflict handling is needed since threads access contiguous addresses. Encoded lengths and minimum values are stored at the beginning, similar to K Cache format, shown in Figure~\ref{fig:v_format}.

\subsection{Complexity Analysis}

This section analyzes the computational and memory complexity of the PackKV. We denote the KV cache dimensions as [\textit{context\_len}, \textit{head\_num}, \textit{head\_dim}], abbreviated as $L, H, D$ respectively. For this analysis, we treat both the truncated block size $N$ (e.g., 64) and the head dimension $D$ (e.g., 128) as fixed constants for a given model.

\textbf{Time Complexity.} The analysis is separated into the incremental compression cost and the recurring decompression cost during decoding.

\textit{1) Compression Pipeline:} The compression of a new $N \times D$ block involves three sequential steps: Quantization, Encode-aware Repacking, and Bit-packing. Since both the block size $N$ and the head dimension $D$ are treated as constants, the time complexity to process one full block is effectively constant, i.e., $O(1)$. Therefore, the amortized compression cost per new token is \textbf{constant time}. 

\textit{2) Decompression and Computation:} A standard attention mechanism performs a matrix-vector multiplication with a time complexity of $O(L \cdot D)$ per attention head. With $D$ being a constant, this simplifies to $O(L)$. 
A two-step approach (decompress then compute) would also have an $O(L)$ complexity but with a large constant factor due to memory write-backs. In contrast, \textbf{PackKV's fused kernel} integrates decompression directly into the computation. While the asymptotic complexity remains $O(L)$, our approach significantly reduces the practical execution time by eliminating the memory bottleneck, leading to substantial throughput improvements.

\textbf{Space Complexity.} PackKV's primary goal is to minimize the memory footprint of the KV cache.

\textit{1) KV Cache Storage:} A standard uncompressed KV cache occupies $M(L \cdot H \cdot D)$ space. Since $D$ is a constant, this scales as $M(L \cdot H)$. With PackKV, this is reduced to  $M(\frac{L \cdot H}{\text{CR}})$, where CR is the compression ratio (e.g., ~15.3x for K cache, ~18.7x for V cache, on average). This makes the memory usage scale much more gracefully with long contexts.

\textit{2) Auxiliary Memory:} PackKV requires a temporary buffer of size [\textit{max\_buffer\_size}, $H, D$], resulting in a space complexity of $M(H \cdot D)$. As both $H$ and $D$ are fixed for a given model, the auxiliary memory requirement is \textbf{constant} and does not scale with the context length $L$,
brings minimum overhead to overall memory usage.

%% file: sections/05_experiment.tex
\section{Evaluation}
\label{sec:evaluation}

\subsection{Experiment Setup}
\label{subsec:exp_setup}

We conduct experiments on two systems: (1) a workstation with AMD Ryzen 7 9800X3D (8 cores), 96 GB RAM, and \textbf{NVIDIA RTX PRO 6000 GPU (98 GB VRAM)}; (2) a multi-node large-scale cluster, each node are equipped with AMD EPYC 7543P (32 cores), 512 GB RAM, and four \textbf{NVIDIA A100 GPUs (40 GB VRAM)}.


Our evaluation metrics are \textbf{Model Accuracy}, \textbf{Compression Ratio}, and \textbf{Decompression Throughput (including computation)}. We selected six models (i.e., Llama2-7B/13B, Llama3.1-8B, DeepSeek-R1-Llama-8B, Ministral-8B-2410, Phi-4) and six benchmarks (i.e., CoQA, GSM8K, MMLU, Winogrande, GPQA (Diamand[zeroshot+n\_shot] as GPQA\_D), SQuAD Completion as SQuAD\_C) for accuracy evaluation. We use KIVI quantization~\cite{liu2024kivi} as the baseline for accuracy and compression ratio. For decompression throughput, we compare our computation-aware decompression+matrix\_vector\_multiplication kernel with original cuBLAS matrix vector multiplication.

Our configuration includes three key parameters: 
\begin{itemize}
    \item \textbf{Relative quantization scale}: scaling factor in $[0, 1]$ where actual scale = $rel\_quant\_scale \times (max\_value - min\_value)$. $rel\_error\_bound = rel\_quant\_scale / 2$.
    \item \textbf{Bit-packing block size}: number of quantized data points sharing encode length and minimum value. 
    \item \textbf{Encode-aware repacking method}: algorithm for repacking quantized KV cache vectors
\end{itemize}

Fixed configurations: \textbf{Maximum buffer size}, \textbf{Truncated block size}. Both are consistent with KIVI.



\subsection{Bit-packing Size Selection}
\begin{figure}[htbp]
    \centering
    \begin{subfigure}[b]{\columnwidth}
        \centering
        \includegraphics[width=0.9\textwidth]{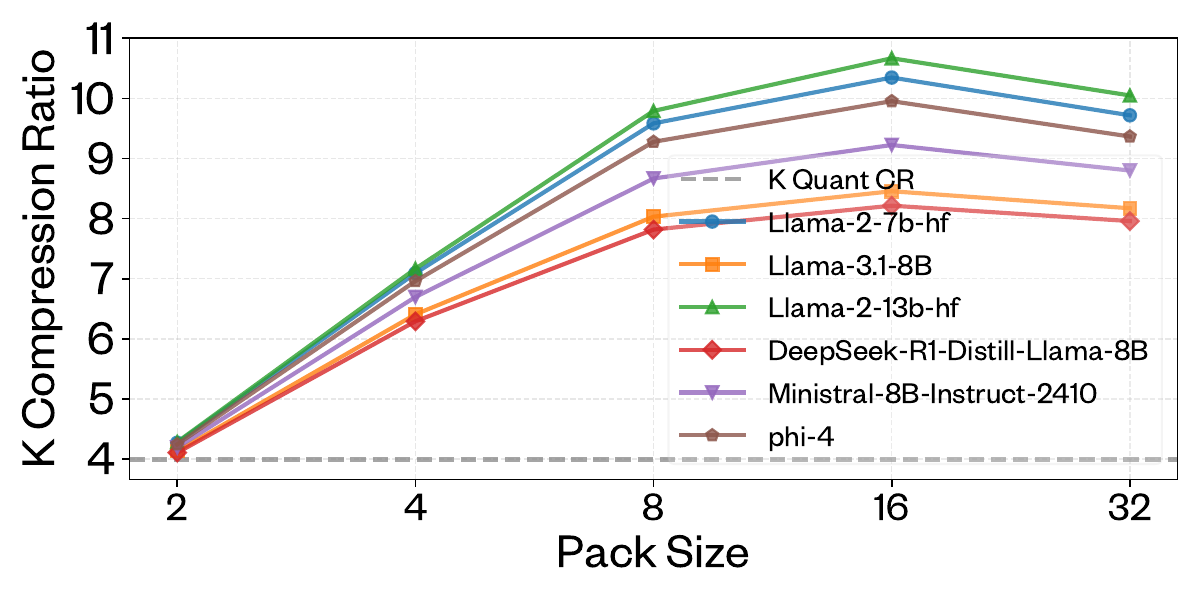}
        \vspace{-0.7em}
        \caption{K Cache Compression Ratio vs Pack Size.}
        \label{fig:pack_size_cr_all_k}
    \end{subfigure}
    \hfill
    \begin{subfigure}[b]{\columnwidth}
        \centering
        \includegraphics[width=0.9\textwidth]{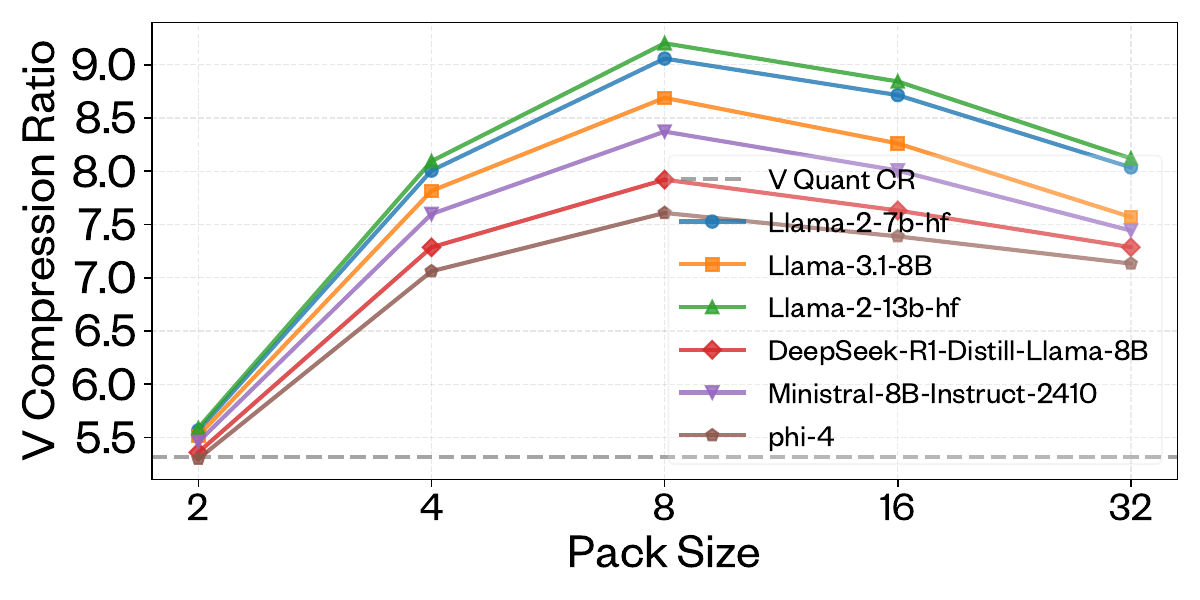}
        \vspace{-0.7em}
        \caption{V Cache Compression Ratio vs Pack Size.}
        \label{fig:pack_size_cr_all_v}
    \end{subfigure}
    \caption{Compression Ratio vs Pack Size with the repacking method set to the optimal choice at each data point.}
    \label{fig:pack_size_cr_all}
    \vspace{-1.2em}
\end{figure}

We evaluate different pack sizes, ranging from 2 to 32, with the K and V relative quantization scales set to 0.1 and 0.2, respectively, which are the appropriate quantization scale settings based on our findings in Table~\ref{tab:turning_points_k} and Table~\ref{tab:turning_points_v}.

\begin{table*}[htbp]
\centering
\caption{Compression ratios with different repacking methods, grouped by cache type (K/V).}
\begin{tabular}{clccccccc}
\toprule
\textbf{Cache} & \textbf{Mode} & \textbf{Llama-2-7B} & \textbf{Llama-3.1-8B} & \textbf{Llama-2-13B} & \textbf{R1-Llama-8B} & \textbf{Ministral-8B} & \textbf{Phi-4} & \textbf{Avg} \\
\midrule
\multirow{3}{*}{K}
    & None   & 10.27 & 7.91 & 10.59 & 7.73 & 8.71 & 9.35 & - \\
    & Greedy & \cdg{10.35 (+0.7\%)} & \cdg{8.45 (+6.9\%)} & \cdg{10.67 (+0.7\%)} & \cdg{8.22 (+6.3\%)} & \cdg{9.22 (+5.9\%)} & \cdg{9.95 (+6.5\%)} & +4.5\% \\
    & Median & 9.93 (-3.4\%) & 7.76 (-1.9\%) & 10.23 (-3.4\%) & 7.55 (-2.2\%) & 8.54 (-2.0\%) & 9.06 (-3.0\%) & -2.7\% \\
\midrule
\multirow{3}{*}{V}
    & None   & 7.29 & 7.22 & 7.37 & 6.89 & 7.01 & 6.64 & - \\
    & Greedy & \cdg{9.06 (+24.3\%)} & \cdg{8.69 (+20.3\%)} & \cdg{9.20 (+24.8\%)} & \cdg{7.92 (+15.0\%)} & \cdg{8.37 (+19.4\%)} & \cdg{7.61 (+14.7\%)} & +19.7\% \\
    & Median & 8.85 (+21.3\%) & 8.56 (+18.5\%) & 9.00 (+22.0\%) & 7.86 (+14.1\%) & 8.24 (+17.5\%) & 7.48 (+12.8\%) & +17.7\% \\
\bottomrule
\end{tabular}

\label{tab:repacking_cr}
\vspace{-2em}
\end{table*}

As shown in Figure~\ref{fig:pack_size_cr_all}, a \textbf{pack size of 8 or 16} yields the best compression ratio in most cases and matches GPU hardware well: with per-value bit-width$<$4, 8 or 16 values occupy exactly 32 or 64 bits, aligning with \texttt{uint32\_t}/\texttt{uint64\_t} and enabling efficient CUDA bitwise shifts and masks.

With optimal repacking, our bit-packing improves compression ratio by \textbf{138.18\% for K} and \textbf{60.05\% for V} over token-wise quantization; both repacking and bit-packing are lossless, so theoretical model accuracy is unchanged compared to quantization while the KV cache size is greatly reduced.

\subsection{Repacking Compression Ratio Improvement}

In this section, we demonstrate the compression ratio improvement achieved by the Encode-aware Repacking method and identify the optimal configuration.

Multiple repacking configurations are supported:
\begin{itemize}
    \item \textbf{Greedy Repacking:} The greedy algorithm~\ref{alg:greedy_repacking}.
    \item \textbf{V Median Repacking:} A simplified repacking method inspired by visualization, proposed here~\ref{sec:sec_v_median}.
    \item \textbf{None:} No repacking is applied.
\end{itemize}



Table~\ref{tab:repacking_cr} compares repacking to token-wise quantization without repacking. Greedy Repacking yields the largest gains (\textbf{K: +4.5\%}, \textbf{V: +19.7\%}); Median Repacking helps mainly on V (\textbf{K: -2.7\%}, \textbf{V: +17.7\%}). For K, bit-packing accounts for \textbf{92.3\%} of the compression-ratio improvement (repacking \textbf{7.7\%}), while for V the contributions are more balanced (repacking \textbf{43.8\%}, bit-packing \textbf{56.2\%}).

Repacking is a lossless transform applied only at compression time.
Attention is permutation-invariant, so the original order does not need to be restored during computation. 
Hence it adds no runtime overhead to the system.

\begin{figure*}[htbp]
    \centering
    \begin{subfigure}[b]{0.32\linewidth}
    \captionsetup{font=scriptsize}
        \includegraphics[width=\linewidth]{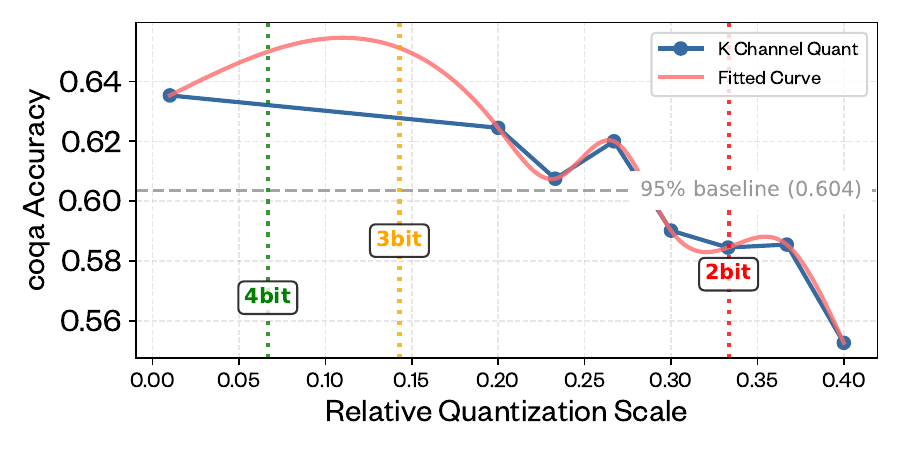}
        \caption{CoQA, Llama2 7B, K Channel Quant(KIVI).}
        \label{fig:accuracy_llama2_7b_k_channel_quant}
    \end{subfigure}
    \begin{subfigure}[b]{0.32\linewidth}
    \captionsetup{font=scriptsize}
        \includegraphics[width=\linewidth]{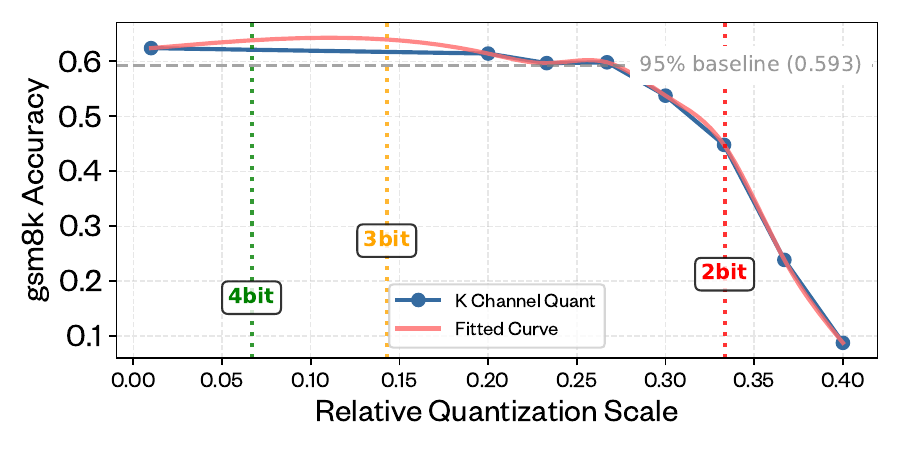}
        \caption{GSM8K, R1 8B(Llama), K Channel Quant(KIVI).}
        \label{fig:accuracy_deepseek_r1_llama_k_channel_quant}
    \end{subfigure}
    \begin{subfigure}[b]{0.32\linewidth}
    \captionsetup{font=scriptsize}
        \includegraphics[width=\linewidth]{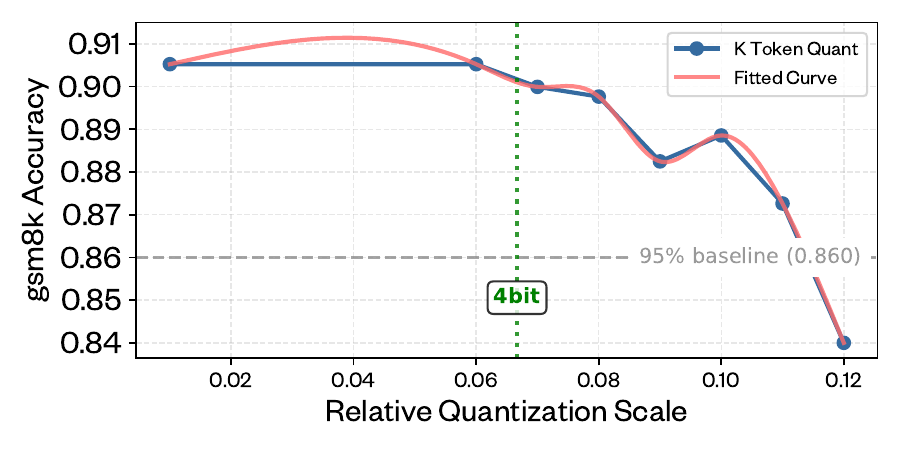}
        \caption{GSM8K, Phi-4, K Token Quant(PackKV-Ours).}
        \label{fig:accuracy_phi_4_k_token_quant}
    \end{subfigure}
    \\
    \begin{subfigure}[b]{0.32\linewidth}
    \captionsetup{font=scriptsize}
        \includegraphics[width=\linewidth]{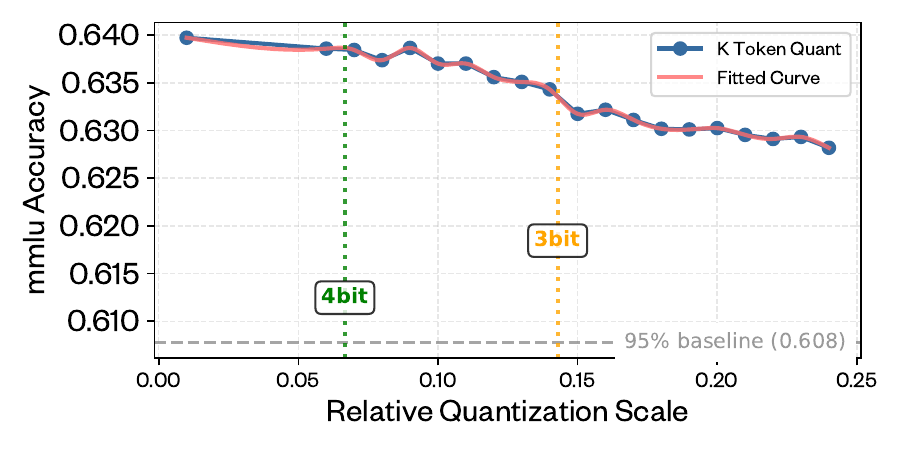}
        \caption{MMLU, Ministral 8B, K Token Quant(PackKV).}
        \label{fig:accuracy_ministral_k_token_quant}
    \end{subfigure}
    \begin{subfigure}[b]{0.32\linewidth}
    \captionsetup{font=scriptsize}
        \includegraphics[width=\linewidth]{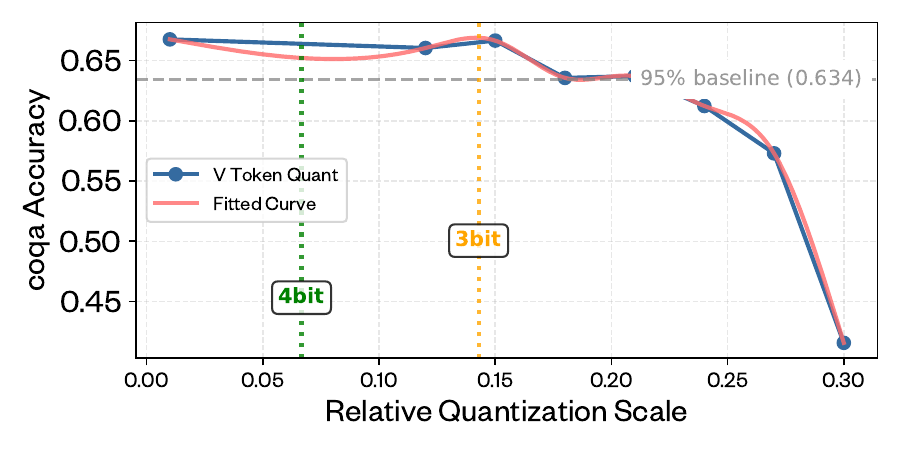}
        \caption{CoQA, Llama2 13B, V Token Quant(KIVI, PackKV).}
        \label{fig:accuracy_llama2_13b_v_token_quant}
    \end{subfigure}
    \begin{subfigure}[b]{0.32\linewidth}
    \captionsetup{font=scriptsize}
        \includegraphics[width=\linewidth]{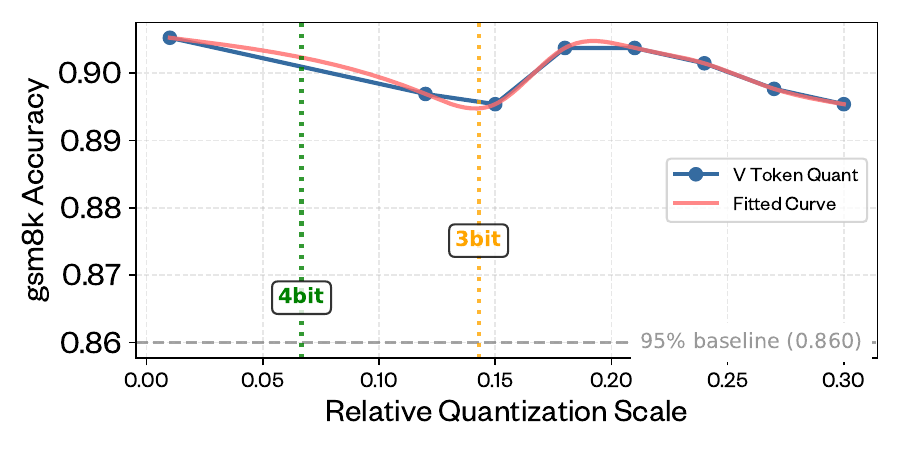}
        \caption{GSM8K, Phi 4, V Token Quant(KIVI, PackKV).}
        \label{fig:accuracy_phi_4_v_token_quant}
    \end{subfigure}
    \caption{Sampled Benchmark accuracies vs Relative quantization scales with the relative quantization scales corresponding to 2,3,4 bit quantization marked. Gray line indicates 95\% of no compression accuracy. Figures show that KIVI(K Channel Quant, V Token Quant) can not stably archive 2 bit quantization while maintaining inference accuracy. }
    \label{fig:accuracy}
\end{figure*}

\begin{table*}[h]
\centering
\scriptsize
\caption{Key Optimal Compression Ratios for KIVI and PackKV(Ours) by Benchmark and Model.
Range denotes the testing relative quantization scale range. PackKV delivers significantly higher compression ratio than KIVI.
}\begin{tabular}{|l|l|l|c|c|c|c|c|c|c|}
\hline
Benchmark & Quant & Range & R1-Llama-8B & Llama-2-13B & Llama-2-7B & Llama-3.1-8B & Phi-4 & Ministral-8B & Avg \\
\hline
CoQA & Channel & [0.01, 0.40] & 4.57 & 4.57 & 4.57 & 4.57 & 6.40 & 6.40 &  \\
 & Token & [0.01, 0.12] & 7.87(+72.1\%) & 10.92(+138.8\%) & 8.44(+84.5\%) & 7.46(+63.3\%) & 10.04(+56.9\%) & 10.54(+64.6\%) & +80.1\% \\
\hline
GPQA\_D & Channel & [0.01, 0.80] & 6.40 & 6.40 & 6.40 & 6.40 & 6.40 & 6.40 &  \\
 & Token & [0.01, 0.24] & 20.80(+225.0\%) & 34.81(+443.8\%) & 33.88(+429.3\%) & 22.06(+244.7\%) & 31.14(+386.6\%) & 26.50(+314.0\%) & +340.6\% \\
\hline
GSM8K & Channel & [0.01, 0.40] & 4.57 & 4.57 & 4.57 & 3.56 & 6.40 & 4.57 &  \\
 & Token & [0.01, 0.12] & 8.91(+94.9\%) & 7.95(+73.9\%) & 7.67(+67.9\%) & 6.52(+83.4\%) & 11.01(+72.0\%) & 8.90(+94.8\%) & +81.2\% \\
\hline
MMLU & Channel & [0.01, 0.80] & 6.40 & 6.40 & 6.40 & 6.40 & 6.40 & 6.40 &  \\
 & Token & [0.01, 0.24] & 20.77(+224.6\%) & 20.42(+219.0\%) & 21.00(+228.1\%) & 22.06(+244.7\%) & 32.36(+405.6\%) & 26.50(+314.0\%) & +272.7\% \\
\hline
SQuAD\_C & Channel & [0.01, 0.40] & 6.40 & 6.40 & 6.40 & 6.40 & 6.40 & 6.40 &  \\
 & Token & [0.01, 0.12] & 9.33(+45.8\%) & 12.65(+97.7\%) & 12.22(+91.0\%) & 9.63(+50.4\%) & 11.84(+85.1\%) & 10.54(+64.6\%) & +72.4\% \\
\hline
Winogrande & Channel & [0.01, 0.40] & 6.40 & 6.40 & 6.40 & 6.40 & 6.40 & 6.40 &  \\
 & Token & [0.01, 0.12] & 9.33(+45.8\%) & 12.65(+97.7\%) & 12.22(+91.0\%) & 9.63(+50.4\%) & 11.84(+85.1\%) & 10.54(+64.6\%) & +72.4\% \\
\hline
Avg & &  & +118.0\% & +178.5\% & +165.3\% & +122.8\% & +181.9\% & +152.8\% & \cdg{+153.2\%} \\
\hline
\end{tabular}

\label{tab:optimal_cr_k}
\vspace{-10pt}
\end{table*}


\subsection{Model Accuracy vs Compression Ratio}

\begin{table*}[h]
\centering
\caption{K cache turning points(5\% accuracy drop) by Benchmarks, Quantization Modes and Models. Range means the relative quantization range we have run benchmarks on. The turning points are the highest relative quantization scales archiving less than 5\% accuracy drop.}
\begin{tabular}{|l|l|l|c|c|c|c|c|c|}
\hline
Benchmark & Quant Mode & Range & R1-Llama-8B & Llama-2-13B & Llama-2-7B & Llama-3.1-8B & Phi-4 & Ministral-8B \\
\hline
CoQA & Channel & [0.01, 0.40] & 0.2724 & 0.3247 & 0.2870 & 0.3045 & 0.3849 & 0.3563 \\
 & Token & [0.01, 0.12] & 0.0903 & 0.1020 & 0.0784 & 0.0787 & 0.0991 & 0.1200 \\
\hline
GPQA\_D & Channel & [0.01, 0.80] & 0.6529 & 0.7914 & 0.8000 & 0.8000 & 0.6794 & 0.7807 \\
 & Token & [0.01, 0.24] & 0.2398 & 0.2400 & 0.2400 & 0.2400 & 0.2342 & 0.2400 \\
\hline
GSM8K & Channel & [0.01, 0.40] & 0.2719 & 0.2516 & 0.2067 & 0.1170 & 0.3977 & 0.3355 \\
 & Token & [0.01, 0.12] & 0.1122 & 0.0682 & 0.0668 & 0.0630 & 0.1142 & 0.0958 \\
\hline
MMLU & Channel & [0.01, 0.80] & 0.3752 & 0.6119 & 0.5940 & 0.3814 & 0.6976 & 0.7717 \\
 & Token & [0.01, 0.24] & 0.2396 & 0.1970 & 0.1874 & 0.2400 & 0.2400 & 0.2400 \\
\hline
SQuAD\_C & Channel & [0.01, 0.40] & 0.3832 & 0.4000 & 0.4000 & 0.4000 & 0.4000 & 0.3852 \\
 & Token & [0.01, 0.12] & 0.1200 & 0.1200 & 0.1200 & 0.1200 & 0.1200 & 0.1200 \\
\hline
Winogrande & Channel & [0.01, 0.40] & 0.4000 & 0.4000 & 0.4000 & 0.4000 & 0.4000 & 0.4000 \\
 & Token & [0.01, 0.12] & 0.1200 & 0.1200 & 0.1200 & 0.1200 & 0.1200 & 0.1200 \\
\hline
\end{tabular}

\label{tab:turning_points_k}
\end{table*}

In this section, we demonstrate that our method can achieve a higher compression ratio than the state-of-the-art quantization method KIVI, while maintaining the same level of accuracy as quantization methods.


\subsubsection{K Accuracy and Compression Ratio Comparison}


We use an adaptive quantization granularity for the K cache than the state-of-the-art quantization method-KIVI.
We compare them at matched accuracy. 
To determine the comparison point, we measure accuracy across relative quantization scales (yielding scatter plots; examples in Figure~\ref{fig:accuracy}) and interpolate to obtain a continuous accuracy–scale curve. We define an acceptable accuracy drop of 5\% (application-dependent; used here only to equalize accuracy) and select the largest relative scale whose loss is $\le$5\%.
These acceptable accuracy turning points are summarized in Table~\ref{tab:turning_points_k}.



Fixing the scales at these turning points, we sweep pack sizes $\{4,8,16\}$ and all repacking methods to maximize compression ratio.
Results are shown in Table~\ref{tab:optimal_cr_k}.
Under the same accuracy, KIVI attains an average K cache compression ratio of \textbf{5.91}, whereas our method achieves \textbf{15.30}, a \textbf{153.2\%} improvement in comparison.

\subsubsection{V Compression Ratio Comparison}

\begin{table*}[h]
\centering
\caption{V cache turning points(5\% accuracy drop) for both PackKV and KIVI by Benchmarks and Models. Range means the relative quantization range we have run benchmarks on. The turning points are the highest relative quantization scales archiving less than 5\% accuracy drop.}
\begin{tabular}{|l|l|l|c|c|c|c|c|c|}
\hline
Benchmark & Quant Mode & Range & R1-Llama-8B & Llama-2-13B & Llama-2-7B & Llama-3.1-8B & Phi-4 & Ministral-8B \\
\hline
CoQA & Token & [0.01, 0.30] & 0.1876 & 0.2160 & 0.1883 & 0.2643 & 0.3000 & 0.2823 \\
\hline
GPQA\_D & Token & [0.01, 0.68] & 0.4763 & 0.6800 & 0.6800 & 0.6800 & 0.6800 & 0.6800 \\
\hline
GSM8K & Token & [0.01, 0.30] & 0.2701 & 0.1246 & 0.1573 & 0.1730 & 0.3000 & 0.3000 \\
\hline
MMLU & Token & [0.01, 0.68] & 0.6287 & 0.5622 & 0.5002 & 0.6298 & 0.6800 & 0.6800 \\
\hline
SQuAD\_C & Token & [0.01, 0.30] & 0.2319 & 0.0415 & 0.0819 & 0.2879 & 0.2828 & 0.1214 \\
\hline
Winogrande & Token & [0.01, 0.30] & 0.3000 & 0.3000 & 0.3000 & 0.3000 & 0.3000 & 0.3000 \\
\hline
\end{tabular}

\label{tab:turning_points_v}
\end{table*}

Unlike the K cache, we apply token-wise quantization to the V cache, following the KIVI quantization approach. This ensures that, given the same quantization settings, the model accuracy theoretically remains unchanged for the V cache. Therefore, to compare the V cache compression ratios, we only need to consider the quantization compression ratio and the encoding compression ratio. First, similar to the K cache experiments, we identify the "turning point," as shown in Table~\ref{tab:turning_points_v}. Next, we compare the compression ratios before and after encoding, with the results presented in Table~\ref{tab:optimal_cr_v}. On average, the V cache compression ratio of \textbf{KIVI quantization is 6.00}, whereas \textbf{our method achieves 18.67}, representing an improvement of \textbf{179.6\%}.

\begin{table*}[h]
\centering
\scriptsize
\caption{V cache optimal compression ratios by benchmark and model (token quantization). Each cell shows cr1/cr2, where cr1 is the KIVI compression ratio and cr2 is the PackKV compression ratio, at matched accuracy levels.}
\begin{tabular}{|l|c|c|c|c|c|c|c|}
\hline
Benchmark & R1-Llama-8B & Llama-2-13B & Llama-2-7B & Llama-3.1-8B & Phi-4 & Ministral-8B & Avg \\
\hline
CoQA & 5.28/7.69(+45.7\%) & 5.32/9.71(+82.4\%) & 5.32/8.69(+63.4\%) & 5.28/10.85(+105.6\%) & 5.29/9.97(+88.4\%) & 5.28/10.86(+105.7\%) & +81.9\% \\
\hline
GPQA\_D & 7.88/21.91(+178.2\%) & 7.98/49.74(+523.7\%) & 7.97/46.61(+484.9\%) & 7.88/36.60(+364.6\%) & 7.90/34.93(+342.1\%) & 7.88/36.65(+365.2\%) & +376.5\% \\
\hline
GSM8K & 5.28/9.85(+86.6\%) & 3.99/6.50(+62.7\%) & 5.32/7.81(+46.8\%) & 5.28/7.99(+51.4\%) & 5.29/9.97(+88.4\%) & 5.28/11.39(+115.8\%) & +75.3\% \\
\hline
MMLU & 7.88/33.86(+329.9\%) & 7.98/43.27(+442.6\%) & 7.97/34.08(+327.7\%) & 7.88/37.13(+371.4\%) & 7.90/34.93(+342.1\%) & 7.88/36.65(+365.2\%) & +363.2\% \\
\hline
SQuAD\_C & 5.28/8.80(+66.8\%) & 3.20/4.09(+28.0\%) & 3.99/5.33(+33.6\%) & 5.28/11.65(+120.7\%) & 5.29/9.58(+81.1\%) & 3.97/5.96(+50.1\%) & +63.4\% \\
\hline
Winogrande & 5.28/10.66(+102.0\%) & 5.32/12.64(+137.5\%) & 5.32/12.38(+132.8\%) & 5.28/12.06(+128.4\%) & 5.29/9.97(+88.4\%) & 5.28/11.39(+115.8\%) & +117.5\% \\
\hline
Avg & +134.9\% & +212.8\% & +181.5\% & +190.4\% & +171.8\% & +186.3\% & \cdg{+179.6\%} \\
\hline
\end{tabular}

\label{tab:optimal_cr_v}
\vspace{-10pt}

\end{table*}

\subsection{Computation Acceleration}


As discussed in Section~\ref{sec:algorithm_design}, during the decoding stage of LLM inference, each iteration requires compressing the KV cache corresponding to a single input token, while decompressing the entire KV cache for the context. This leads to a significant imbalance between compression and decompression workloads. Consequently, our evaluation focuses on decompression throughput and the resulting acceleration of inference computation.


\begin{figure}[]
    \centering
    \begin{subfigure}[b]{0.49\columnwidth}
        \includegraphics[width=\textwidth]{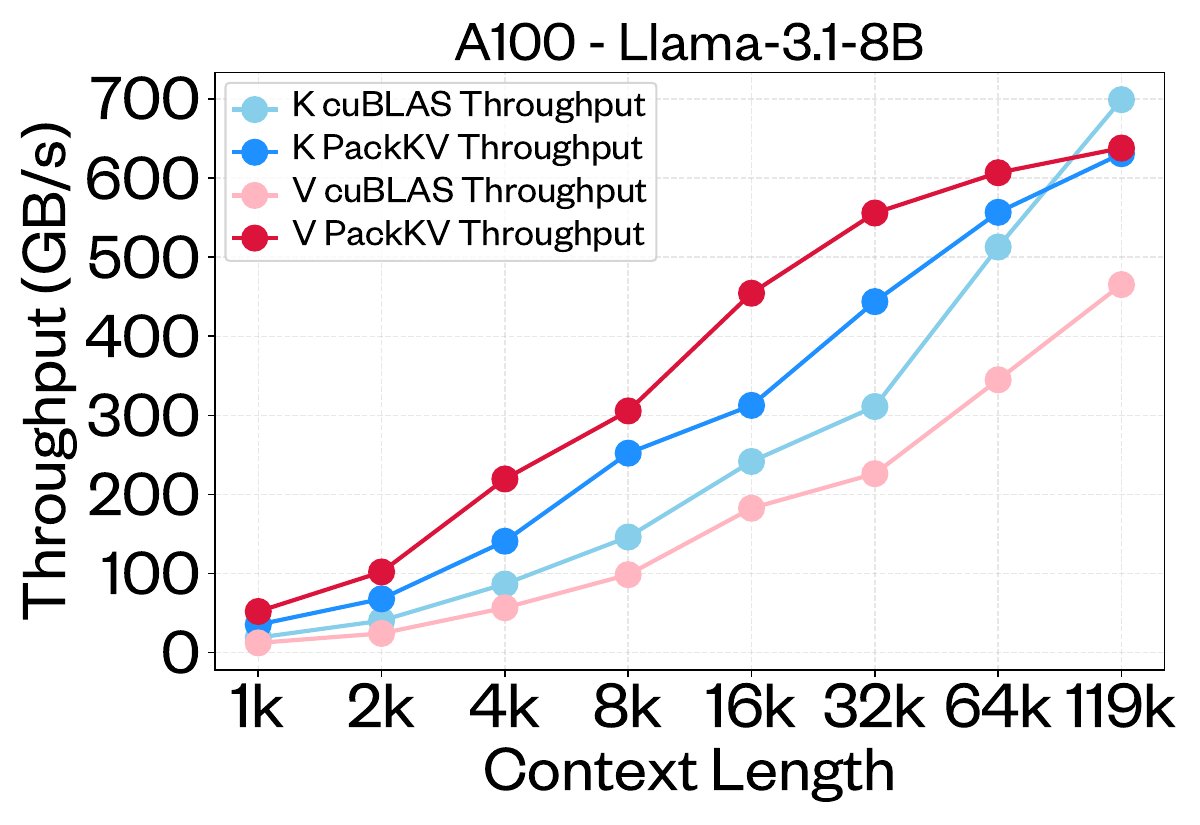}
        \label{fig:a100_detailed_llama}
    \end{subfigure}
    \hfill
    \begin{subfigure}[b]{0.49\columnwidth}
        \includegraphics[width=\textwidth]{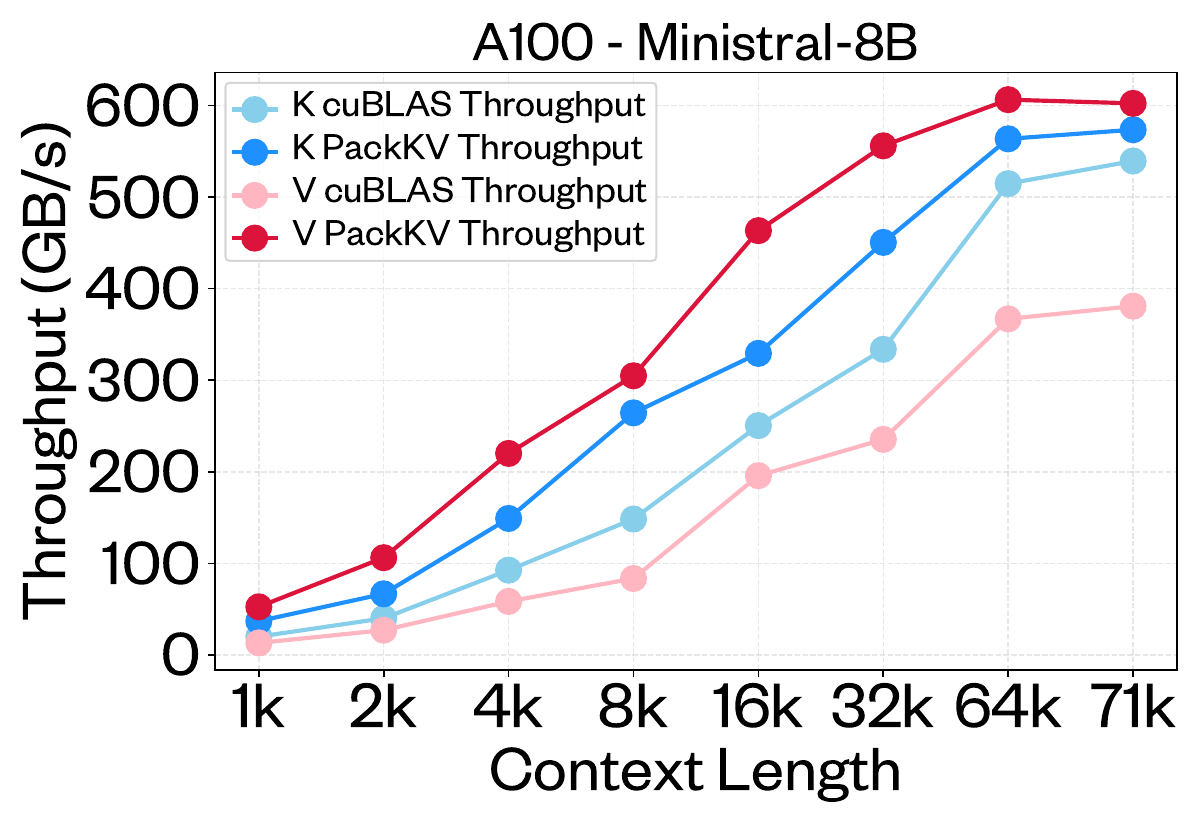}
        \label{fig:a100_detailed_ministral}
    \end{subfigure}
    \vspace{-2em}
    \caption{Throughput comparison between cuBLAS MatVec Multiplicatin kernel throughput and PackKV decompression+matvec multiplicaion kernel throughput on A100 GPU across different models and cache components.}
    \label{fig:a100_results}
\vspace{-1em}
\end{figure}

We select the Mistral-8B and Llama 3.1 8B models, both offer the longest context length (128k) among our chosen models. Contexts are constructed using the compression configurations: K relative quantization scale = 0.1, V relative quantization = 0.2, bit packing method = None, and pack size = 16. KV cache data is collected during model inference. We then perform matrix-vector multiplication on the collected data, comparing the kernel execution time for cuBLAS (in PyTorch) matrix-vector multiplication on the original data with our fusion computation+decompression kernel time on the compressed data.



\begin{figure}[]
    \centering
    \begin{subfigure}[b]{0.49\columnwidth}
        \includegraphics[width=\textwidth]{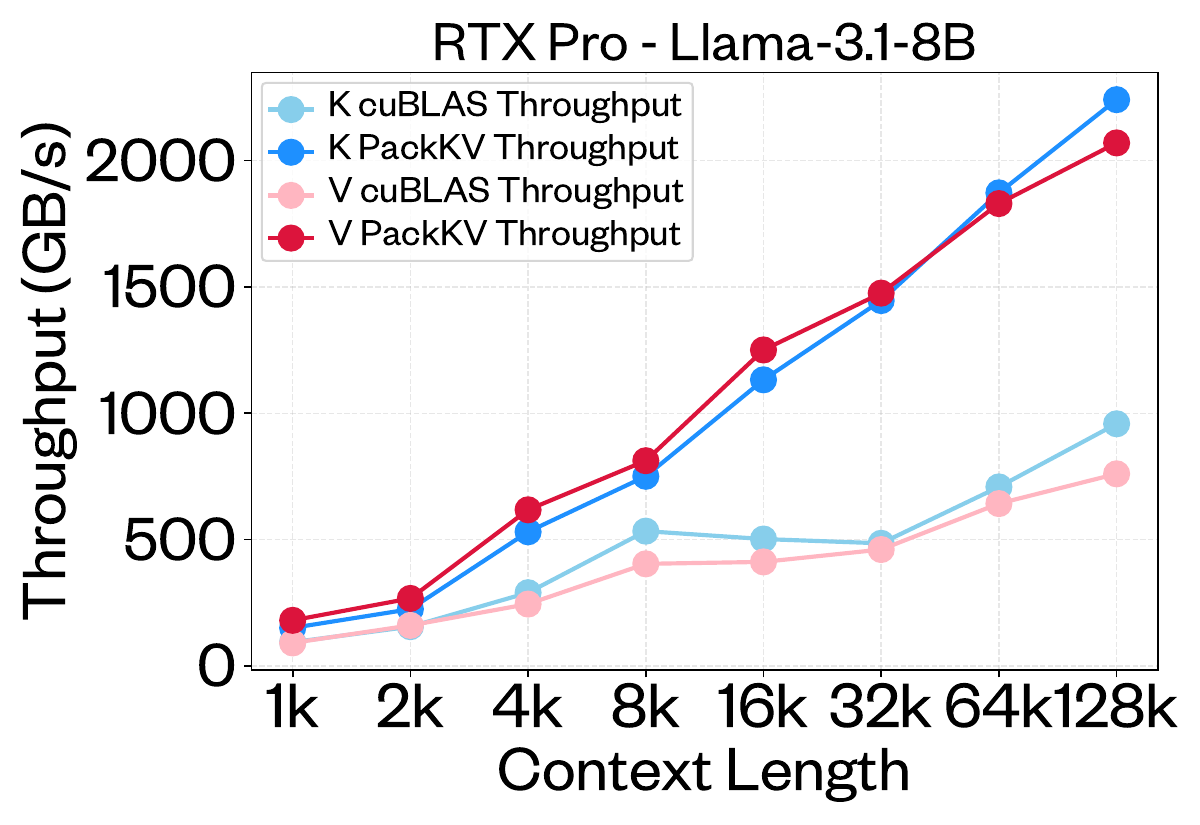}
        \label{fig:rtx_detailed_llama}
    \end{subfigure}
    \hfill
    \begin{subfigure}[b]{0.49\columnwidth}
        \includegraphics[width=\textwidth]{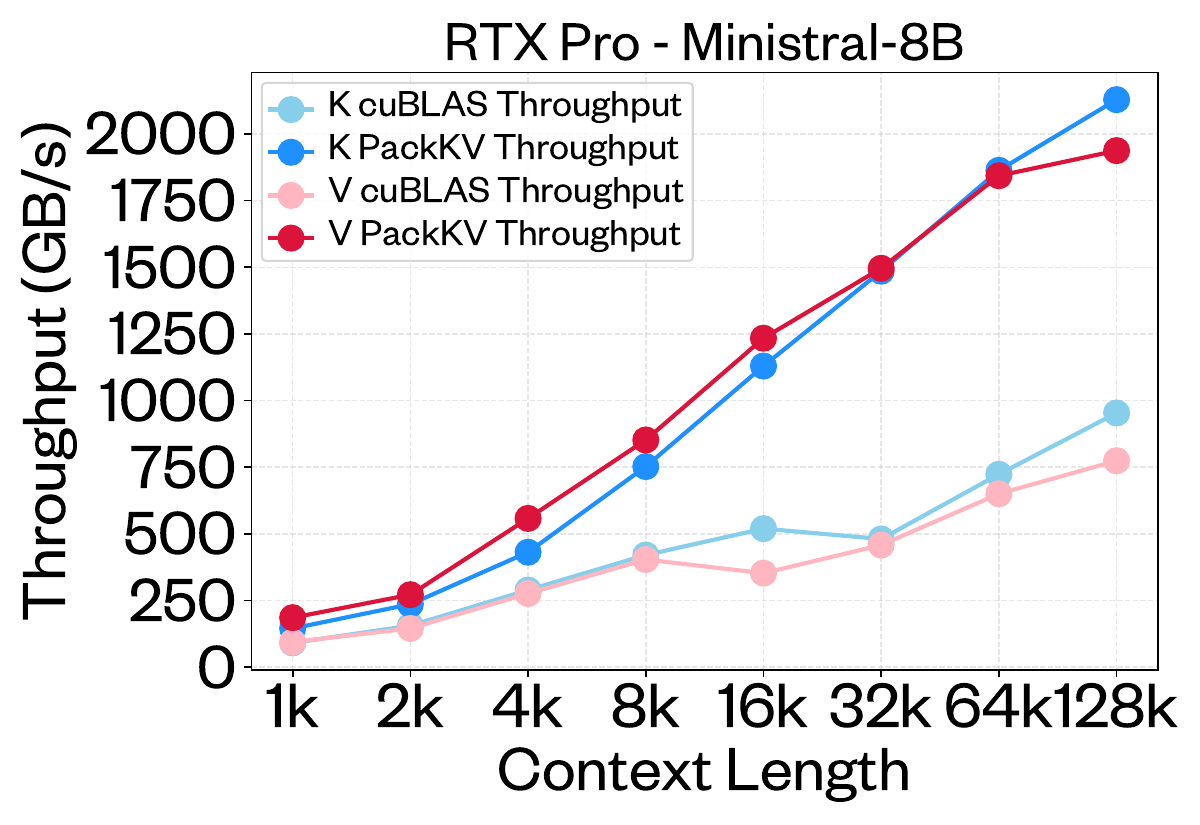}
        \label{fig:rtx_detailed_ministral}
    \end{subfigure}
    \vspace{-2em}
    \caption{Throughput comparison between cuBLAS MatVec Multiplicatin kernel throughput and PackKV decompression+matvec multiplicaion kernel throughput on RTX pro 6000 GPU across different models and cache components.}
    \label{fig:rtx_results}
\vspace{-1.5em}
\end{figure}

On both workstation- and cluster-GPUs, our method markedly boosts matrix–vector throughput: on A100 shown in figure~\ref{fig:a100_results}, +45.1\% for K and +193.5\% for V; on RTX Pro 6000 shown in figure~\ref{fig:rtx_results}, +106.2\% for K and +149.8\% for V, with kernel time reduced accordingly.


In conclusion, our method improved matrix multiplication throughput by \textbf{75.6\% for k and 171.6\% for v}, on average. We also measured the proportion of time spent on the None matrix-vector multiplication kernel in our implementation, which accounts for 25.3\% of the total overhead. This overhead can be further reduced by integrating the partial de-quantization process into our fusion decompression+matrix-vector multiplication kernel.

\subsection{Scalability Evaluation}

\begin{figure}[]
    \centering
    \begin{subfigure}[b]{0.48\columnwidth}
        \includegraphics[width=\textwidth]{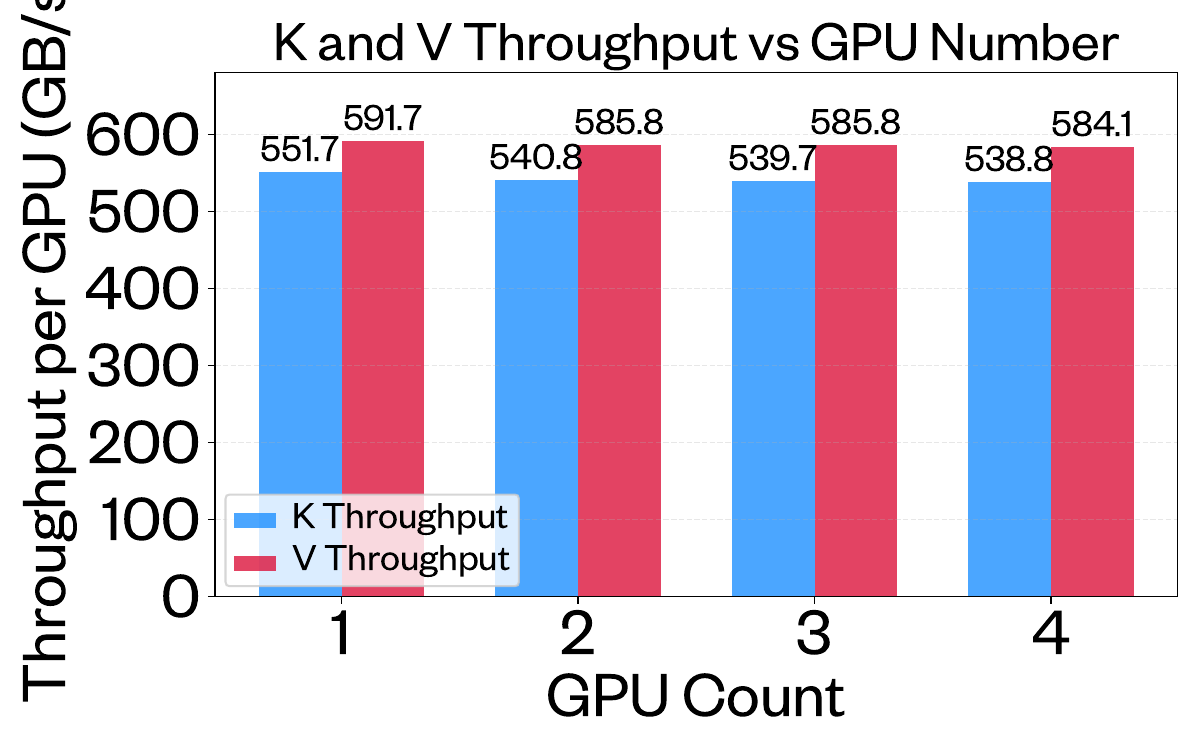}
        \caption{Per-GPU throughput.}
    \end{subfigure}
    \hfill
    \begin{subfigure}[b]{0.48\columnwidth}
        \includegraphics[width=\textwidth]{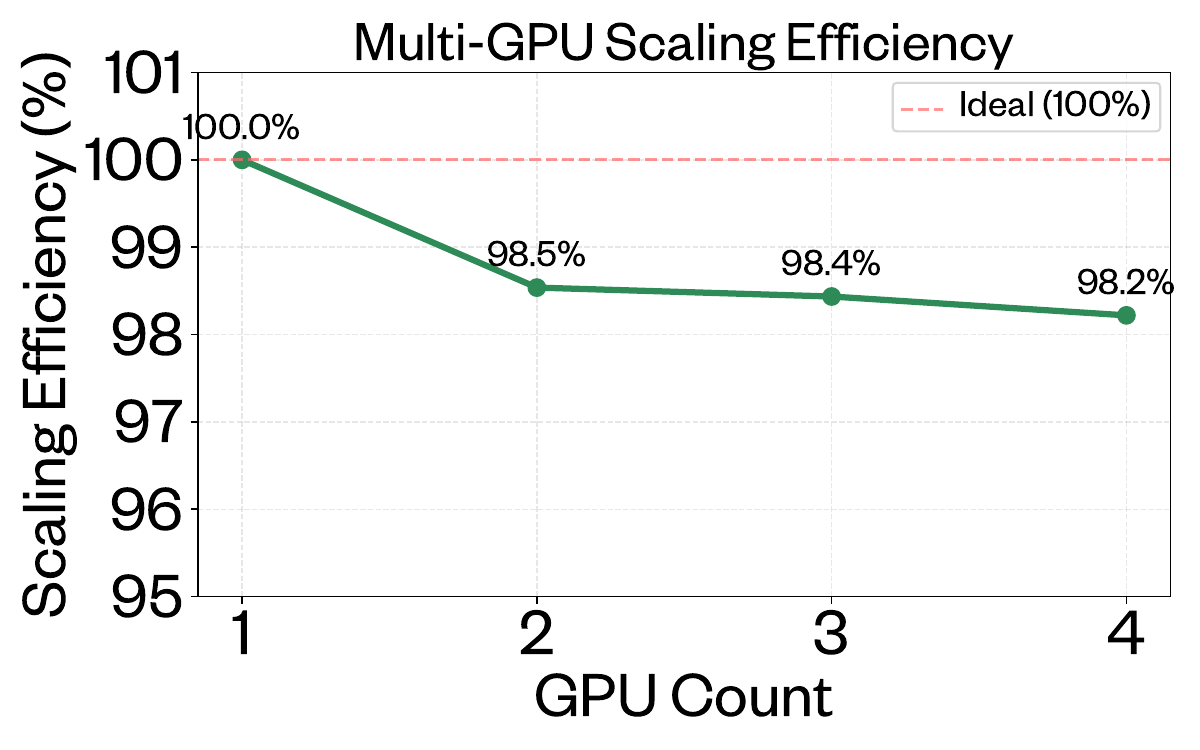}
        \caption{Scaling efficiency.}
    \end{subfigure}
    \caption{Scalability of PackKV. Evaluate PackKV performance with different number of A100 GPUs.}
    \label{fig:throughput_multi_gpu}
\vspace{-1.3em}
\end{figure}

PackKV is theoretically unaffected by multi-GPU deployment, we conduct experiments to empirically evaluate  running multiple instances on multiple GPUs.
Specifically, we use the Llama-3.1-8B model with a context length of 32k, and design 100 throughput test cases. The tasks are evenly distributed across 1, 2, 3, or 4 A100 GPUs on a single node. We then collect the throughput results and calculate the per-GPU throughput to assess potential interference among instances. As shown in Figure~\ref{fig:throughput_multi_gpu}, the per-GPU throughput remains consistent regardless of the number of GPUs used within a single node, indicating that there is minimum interference between GPUs. 
Based on these results, we can further infer that there would also be no interference between nodes. Thus, our framework achieves near-perfect scaling.

%% file: sections/06_conclusion.tex
\section{Conclusion and Future Work}
\label{sec:conclusion}

In this paper, we presented \textbf{PackKV}, a generic and efficient lossy compression framework for managing the KV cache in LLM inference, particularly optimized for long-context generation. 
With no accuracy loss relative to state-of-the-art quantization baselines, PackKV
reduces KV cache memory footprint by 15.30$\times$(K) and 18.67$\times$(V) on average, a \textbf{153.2\%}(K) and \textbf{179.6\%}(V) improvement compared to previous quantization-only solution.
While accelerate execution throughput by \textbf{75.7\%} (K) and \textbf{171.7\%} (V) on average compared to cuBLAS matrix–vector kernels on A100 and RTX Pro 6000 GPUs, respectively,
using less memory bandwidth and incurring effectively zero decompression overhead.

In the future, we plan to explore more advanced encoding methods, develop improved repacking algorithms, and further optimize the computation-aware decompression kernel
to enhance both compression ratio and throughput. 
We will also extend PackKV to a broader set of LLMs and GPUs.

\newpage